# Structuring data analysis projects in the Open Science era with Kerblam!

**Luca Visentin**[*1] , **Luca Munaron**[*2] , **Federico Alessandro Ruffinatti**[*3]

*: Department of Life Sciences and Systems Biology, University of Turin, Turin, Italy

1: L.V., Email: (Corresponding Author) luca.visentin@unito.it , 2: L.M., Email: luca.munaron@unito.it , 3: F.A.R., Email: federicoalessandro.ruffinatti@unito.it

## Abstract

Structuring data analysis projects, that is, defining the layout of files and folders needed to analyze data using existing tools and novel code, largely follows personal preferences. In this work, we look at the structure of several data analysis project templates and find little structural overlap. We highlight the parts that are similar between them, and propose guiding principles to keep in mind when one wishes to create a new data analysis project. Finally, we present Kerblam!, a project management tool that can expedite project data management, execution of workflow managers, and sharing of the resulting workflow and analysis outputs. We hope that, by following these principles and using Kerblam!, the landscape of data analysis projects can become more transparent, understandable, and ultimately useful to the wider community.

## 1 Introduction

Data analysis is a key step in every scientific experiment. In numerical-data-centric fields, data analysis is, in essence, a series of computational steps in which input data is processed by software to produce some output. Usually, the ultimate goal of such analyses is to create secondary data for human interpretation in order to produce knowledge and insight on the raw input data. These manipulations can involve downloading input data on local storage, creating workflows and novel software—also saved locally—and running the analysis on local or remote ("cloud") hardware.

In this article, we will use the phrase "data analysis project structure" to refer to the way data analysis projects are organized on the actual file system, including the structure of folders on disk, the places where data, code, and workflows are stored, and the format in which the project is shared with the public. Unfortunately, such structures can vary a lot from one researcher to another, making it hard for the public to inspect and understand them.

With the Open Science movement gaining more and more traction in the recent years [1], there is a growing need to standardize how routine data analysis projects are structured and carried out. Notably, even if originally thought to provide guidelines for the management of data, FAIR principles have recently been extended also to other contexts, such as that of software [2]. In this view, by making data analyses more transparent and intelligible, the standardization of project structure complies with the FAIR principles' call for more Findable, Accessible, Interoperable, and Reusable research objects [3]. Consistently, efforts are being made from many parts to make reproducible pipelines easier to be created and executed by the wider public—for example by leveraging meth-ods such as containerization. However, while new tools and technologies offer unprecedented opportunities to make the whole process of data analysis increasingly transparent and reproducible, their usage still takes time and effort, as well as expertise and sensibility to the issue of standardization and reproducibility by the experimenter.

In this work, we inspect the structure of many data science and data analysis project templates currently available online. Then, we outline best practices and considerations to take into account when thinking about structuring data analysis projects. Following these principles, we propose a simple, lightweight and extensible project structure that fits many needs and is in line with the projects already present in the ecosystem, thus providing a certain level of standardization. Finally, we introduce Kerblam!, a new tool that can be used to work in projects with this standard structure, taking care of common tasks like data retrieval and cleanup, as well as workflow management and containerization support. This could ultimately benefit the scientific community by making others' work easier to understand and reproduce, for example during the peer review process.

## 2 Data Collection

To fetch the structure of the most common data analysis projects, we ran two GitHub searches: one for the keywords *cookiecutter* and *data* (`cookiecutter` is a Python package that allows users to create, or "cut", new projects from templates) and the other for the much more generic keywords *project* and *template*.

We downloaded the top $50$ such repositories from each search sorted by GitHub stars, as a proxy for popularity and adoption rate. For each downloaded project, we either cut it with the `cookiecutter` Python package or used it as-is (for non-cookiecutter templates). Of these $100$ repositories, $87$ could ultimately be successfully cut and parsed, and were therefore considered. All files and folders from the resulting projects were then listed and compiled into a frequency graph.

Some housekeeping files (like the `.git` directory and all its content) were stripped from the final search results as they were deemed not relevant to the project as a whole. For example, `.gitkeep` files—which are commonly used to commit empty directories to version control—were excluded from the final figure. Finally, only files present in at least three or more templates were retained for plotting.

The analysis was performed with the latest commits of all the considered repositories as of the 12th of July, 2024. The only exception was the "drivendataorg/cookiecutter-data-science" repository, for which we fetched version `1.0` due to non-standard parsing requirements of the latest commit.

The code for the analysis is available online. See the "Code and Data availability" section for more information.

## 3 Data Interpretation

The choice of how to structure projects is an issue universally shared by anyone who performs data analysis. This results in a plethora of different tools, folder hierarchies, accepted practices, and customs. To explore the most common project-structuring prac-tices, we inspected $87$ different project templates available on GitHub and produced a frequency graph of shared files and folders visible in Figure 1.

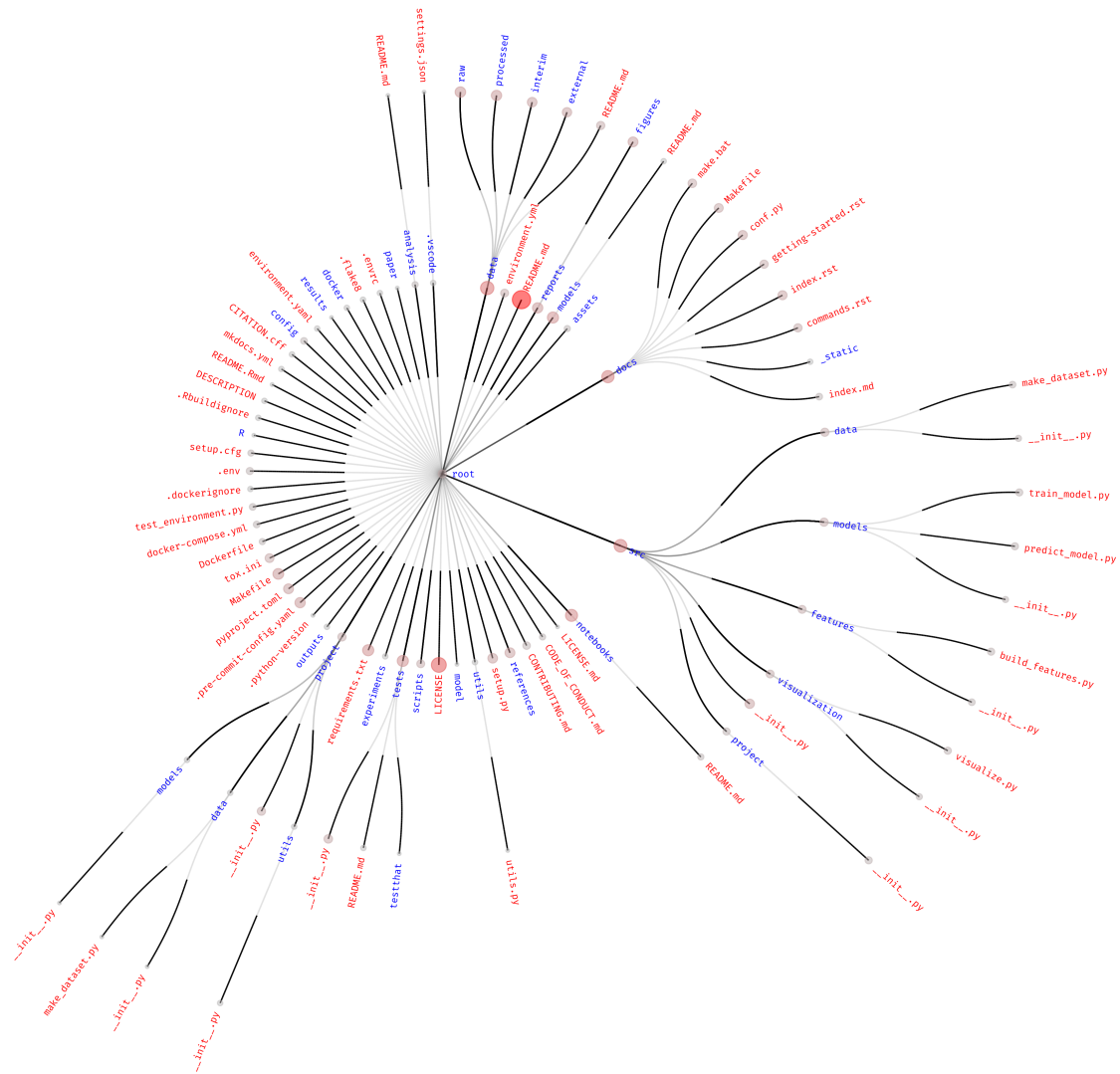

Figure 1: Frequency graph of the structure of the $87$ most starred data analysis project templates, as retrieved from GitHub. Only files present in at least three or more templates are shown. The size and color intensity of the circle at the tip of each link is proportional to the frequency with which this file or folder is found in different project templates. Red text represents files, while blue text represents folders. The central dot of the root node was assigned an arbitrary size.

By looking at this figure, we can point out common patterns in project structure. It must be highlighted, however, that templates influence each other. For example, many Python data science project templates seem to be modified versions of [drivendataorg/cookiecutter-data-science](), which has a very high number of stars and is therefore probably popular with the community.

In any case, the two most highly found files are the `README.md` file (with a frequency of $\frac{77}{87} \simeq 0.89$) and the `LICENSE` and `LICENSE.md` files ($\frac{46+3}{87} \simeq 0.56$).

The `pyproject.toml` file at the top level of the repository—marking the project as a Python package—is also prevalent ($\frac{16}{87} \simeq 0.18$). This is potentially due to the popular

"cookiecutter-data-science" template mentioned before, also highlighting how projects following this template are intimately linked with the usage of Python, potentially exclusively. The predominance of Python-based projects is also noticeable by the presence of `requirements.txt` (a file usually used to store Python's package dependencies), `setup.py`, and `setup.cfg` (now obsolete versions of the `pyproject.toml` file, used to configure Python's build system). The `project` folder at the top level of the templates is most likely the Python package (represented by the `__init__.py` file) that the `pyproject.toml` file refers to (the name "project" is artificial, deriving from the default way that cookiecutter templates were cut).

The presence of files related to the R programming language (the `R` directory, `.Rbuildignore`, `README.Rmd`) reflect its usage in the data analysis field, although at a lower frequency than that of Python. The relatively low prevalence of the R programming language could be due to biases introduced by the search queries, or to the overwhelming popularity of Python project templates, also in the light of the fact that the cookiecutter utility itself is written in Python.

Community-relevant files such as `CONTRIBUTING.md` ($\frac{8}{87} \simeq 0.09$) and `CODE_OF_CONDUCT.md` ($\frac{5}{87} \simeq 0.06$) show little prevalence in templates. This is also true for the `CITATION.cff` ($\frac{4}{87} \simeq 0.05$) file, useful for machine-readable citation data.

The `src` ($\frac{31}{87} \simeq 0.36$), `data` ($\frac{35}{87} \simeq 0.40$), and `docs` ($\frac{28}{87} \simeq 0.32$) folders are very highly represented, containing code, data, and project documentation, respectively. In particular, the `data` directory contains with a high frequency the `raw`, `processed`, `interim`, and `external` folders to host the different data types—input, output, intermediate, and third party—according to the structure promoted by the "cookiecutter-data-science" template. The prevalence of these sub-folders, however, is lower than the frequency of `data` itself, which means that the presence of `data` folder is not uniquely due to that specific template. Interestingly, other templates include `data` in the `src` folder, mixing it with analysis code. Other common folders present in the `src` directory are also the ones promoted by "cookiecutter-data-science", but again, as already noted for `data`, their occurrence is lower than that of the parent folder, indicating that many different templates adopt `src` as a folder name.

Docker-related files are present, mostly in the top-level of the project: `Dockerfile` ($\frac{5}{87} \simeq 0.06$), `.dockerignore` ($\frac{4}{87} \simeq 0.05$), and `docker-compose.yml` or `yaml` ($\frac{6+1}{87} \simeq 0.08$). Docker-related files and folders are also present with sub-threshold frequencies in many other forms, often as directories with multiple Dockerfiles in different folders. The presence of the `docker-compose.yml` file and docker subdirectories could be indicative of a common need to manage multiple execution environments—that work together in the case of Docker Compose—throughout the analysis.

The sparse usage of many tools can be appreciated by the amount of unique files and folders across all templates. Out of $4195$ different files and directories considered by this approach, the vast majority ($3908$, or $93.16\%$) are present in only one template. Looking at directories only, $783$ are unique over $864$ total ($90.63\%$). This figure might be inflated due to the presence of some compiled libraries, files, and Git objects that are included in the analysis and not correctly removed by our filtering. However, we argue that this overwhelmingly high uniqueness would not be significantly affected by manual filtering.

The tiny overlap between templates reflects the fact that project structure is, by its nature, a matter of personal preference. At the same time, Figure 1 confirms that the core structure of the repositories tends to be similar. This is potentially due to both the epistemic need to share one's own work with others and technical requirements of research tools, which cause the adoption of community standards either by choice (in the former case) or imposition (in the latter). For instance, the high presence of the `README.md` file is a community standard that is broadly shared by the majority of software developers, users, and researchers alike. This adoption is purely due to practical reasons: specifically the need to share the description of the work with others in an obvious ("please read me"), logical (in the topmost layer of the project layout), and predictable (i.e., used by the wider community) way.

Borrowing a term from genetics, the `README` file can be thought to be a "housekeeping" file: without it, the usefulness of a project is severely impaired. In this regard, another possible housekeeping candidate is the `LICENSE` file. It is essential for collaborating with the community in the Open Source paradigm, and thus commonly found in many software packages. The concession for code reuse is also essential in data analysis projects, both to allow reproducers to replay the initial work and for other researchers to build on top of previous knowledge. Incidentally, the common presence of the `LICENSE` file in the *template* of a project is interesting. This could be either due to apathy toward licensing issues, leading to picking a "default license" without particular considerations, or a general feeling in individuals that one particular license fits their projects across the board.

A potentially new housekeeping file that is however not yet commonly found is the `CITATION.cff` file. This file contains machine-readable citation metadata, which could be used by both human and machine users to obtain such information, potentially automatically.

## 4 Intervention

### 4.1 Design Principles

The observations made above can all be taken into consideration when designing a more broadly applicable project template that may be used in a variety of contexts. To this end, it is helpful to conceptualize some core guiding principles that should be followed by all data analysis projects, in particular under the Open Science paradigm.

As data analysis projects often involve writing new software, a data analysis project structure requires support for both *data analysis* proper and *software development*. The methods of software development fall outside the scope of this work, but some concepts are useful in the context of data analysis, particularly for *ad hoc* data analysis. For instance, many programming languages require specific folder layouts in order to create self-contained, distributed software. Take, for instance, the Python programming language: to create a Python package, a specific project layout has to be followed [4].

This is visible also in Figure 1, with the presence of the "project" folder, and many files specific to Python packages, crucially, in the locations required by Python build backends. Something similar occurs for many programming languages, such as R [5] and Rust [6], among others.

However, a researcher might not want to create self-contained, distributed software. Languages like Python and R can interpret and execute single-file scripts which achieve some goal (i.e., "scripting"). As scripting is so fast, convenient, and easy to do, it is the most common method of doing data analysis. Scripting provides much flexibility during the development process, but this typically exacerbates the fragmentation of project structures. In particular, the environment of execution now becomes much more relevant: which packages are installed and at which versions, the order that the scripts were read and executed and, potentially, even the order of *which lines* are (manually) run become important to the success of the overall analysis.

This increased flexibility is obviously useful for the research process, which requires the ability to change quickly in order to adapt to new findings, especially during hypothesis-generating "exploratory" research. The principles presented here aim to retain this essential requirement of adaptability, but, at the same time, push for increased standardization of methods, avoiding the most common and dangerous pitfalls one can encounter during the process of data analysis.

*1. Use a version control system.* At its core, software is a collection of text files, and this includes data analysis software. While working on code, it is important to record the differences between the different versions of these files. This is very useful, especially during the research process, to "retrace our steps" or to attempt new methodologies without the fear of losing any previous work. Such records are also useful as provenance information, and potentially as proof of authorship, similarly to what a laboratory notebook does for a "wet-lab" experimental researcher.

There is consolidated software that can be used as a version control system. The overwhelming majority of projects use `git` to this aim, although others exist. Platforms that integrate `git` such as GitHub ([github.com](github.com)) and GitLab ([gitlab.com](gitlab.com)) are increasingly used for data analysis both as a collaboration tool during the project and a sharing platform afterwards.

The first principle should therefore be this: **use a version control system**, such as `git`.

A few practical observations stem from this principle:
- version control encourages good development practices, such as atomic commits, meaningful commit messages, and more, reducing the amount of mistakes made while programming and increasing efficiency by making debugging easier;
- version control discourages the upload of very large (binary) files, so input and output data cannot be efficiently shared through such a system, incentivizing the deposit of data in online archives and—by extension—favoring the FAIR-ness of the manipulated data objects;
- code collaboration and collaboration techniques (such as "GitHub Flow" or "trunk based development" [7,8]) can be useful to promote a more efficient development workflow in data analysis disciplines such as bioinformatics, especially in mid- to large- research groups;
- the core unit of a project should be a code repository, containing everything related to that project—from code, to documentation, to configuration.

The usage of a version control system has implications also for FAIR-ness. Leveraging remote platforms can be fundamental for both Findability and Accessibility. Integrations of platforms like GitHub with archives like Zenodo, for instance, allow developers to easily archive for long-term preservation their data analysis code, promoting Accessibility, Findability, and Reusability.

*2. Documentation is essential.* When working on a data analysis project, documentation is important for both the experimenter themselves and any possible external users. Through ideal documentation, the rationale, the process, and potentially the result of the analysis are presented to the user, together with practical steps on how to *actually* reproduce the work. As with all other aspects of data analysis, documentation takes many different forms, but is the most difficult thing to standardize for one simple reason: documentation is written by humans for human consumption. Documentation is therefore allowed high flexibility in structure, content, form, and delivery method.

Even though rigid standardization is impossible, some guidelines on how to write effective documentation can still be drawn, oftentimes from best practices in the much wider world of Open Source software. We have already highlighted the fundamental role of the `README` file and its very wide adoption. This file contains high-level information about the project and is usually the first—and perhaps only—documentation that all users encounter and read. It is therefore essential that core aspects of the project are delivered through the `README` file, such as:
- the aim of the project, in clear, accessible language;
- the methods used to achieve such aim (and/or a link to further reading material);
- a guide on how to run the analysis on the user's machine, potentially including information on hardware requirements, software requirements, container deployment methods, and every information a human reproducer might need to execute the analysis;
- in an Open Science mindset, including information on how to collaborate on the project and contact information of the authors is also desirable.

Other aspects of the project, such as a list of contributors, might also be included in the `README` file. The `README` file may also be named `DESCRIPTION`, although `README` is a much more widely accepted standard.

Additional documentation can be added to the project in many forms (also see Figure 1). A common documentation file is the `CONTRIBUTING` file, with information on how to contribute to the project, on how authorship of eventual publications will be assigned, and other community-level information. The `CODE_OF_CONDUCT` file contains guidelines and policy on how the project is managed, the expected conduct of project members, and potentially how arising issues between project members are dealt with. Such a file can be important to either projects open to collaboration from the public or large consortium-level projects. Another important documentation file in the Open Source commu-nity is the `CHANGELOG` file. It contains information about how the project changed over time and its salient milestones. For data analysis, it could be used to inform collaborators of important changes in the codebase, methodology or any other news that might be important to announce and record. Additionally, together with the commit history, `CHANGELOG` files can be useful to track the provenance of the analysis, as we have already mentioned.

A common place to store documentation is the top-level of the project repository, but some templates use the `docs` folder, also coming from guidelines used in the Python community (to use tools such as Sphinx [9]).

We can conclude by reiterating that the second principle states that **documentation is essential**.

*3. Be logical, obvious, and predictable.* When a project layout is logical, obvious, and predictable, a human user can easily and quickly understand and interact with it.

To be *logical*, a layout should categorize files based on their content, and logically arrange them following such categories. To be *obvious*, this categorization should make sense at a glance, even to non-experts. For instance, a folder named "scripts" should contain scripts (to be obvious) and only scripts (to be logical). To be *predictable*, a layout should adhere to community standards, so that it "looks" similar to other projects. This creates minimal friction when a user first encounters the project and desires to interact with it.

This principle is present also in other aspects of project structure other than layout. For instance, the structure of documentation can also benefit from the same principles, but in a different context: logically arranged, obvious in structure, and similar to other projects.

This might be the most difficult principle to follow, as it largely depends on the community as a whole. For this reason, we hope that the analysis shown above, especially Figure 1, and our proposed minimal structure (presented in the next sections) will be useful as guides to effectively implement this principle.

We can summarize this third principle like this: **be logical, obvious, and predictable**.

*4. Promote (easy) reproducibility.* Scientific Reproducibility has been and still is a central issue, particularly in the field of biomedical research [10,11]. Scientific software developers hold the crucial responsibility towards the scientific community of creating reproducible data analysis software.

"Reproducibility" can be understood as the ability of a third-party user to understand the research issue investigated by the project, how it was addressed, and practically execute the analysis proper again to obtain a hopefully similar and ideally identical result as the original author(s). This has twofold benefits: a reproducible analysis evokes more confidence in those that read and review it, and it makes it much easier to repurpose the analysis to similar data in the future.

In the modern era, scientists are equipped with powerful tools to enable reproducibility, such as containerization, virtualization, etc. While a discussion on how reproducibility can be achieved eludes the scope of this article, the project layout can promote it, especially when all other principles presented here are respected. This increased adoption can be promoted by including obvious and easily implementable reproducibility methods right in the layout of the project.

Workflow managers, like Nextflow [12], Snakemake [13], and the Common Workflow Language (CWL) [14] are key tools to enable reproducibility. They allow a researcher to describe in detail the workflow used, from input files to final output, offloading the burden of execution to the workflow manager. This allows greater transparency in the methodology used, and even makes reproducibility a possibility in more complex data analysis scenarios. Additionally, some workflow managers are structured to promote reusability of the analysis code, even on different architectures or in high performance computing environments [14].

We can conclude this section by stating the fourth and last principle: **be (easily) reproducible**.

## 4.2 Kerblam!

We have designed a very simple but powerful and flexible project layout together with a project management tool aiming at upholding the principles outlined in the previous section. We named this tool "Kerblam!".

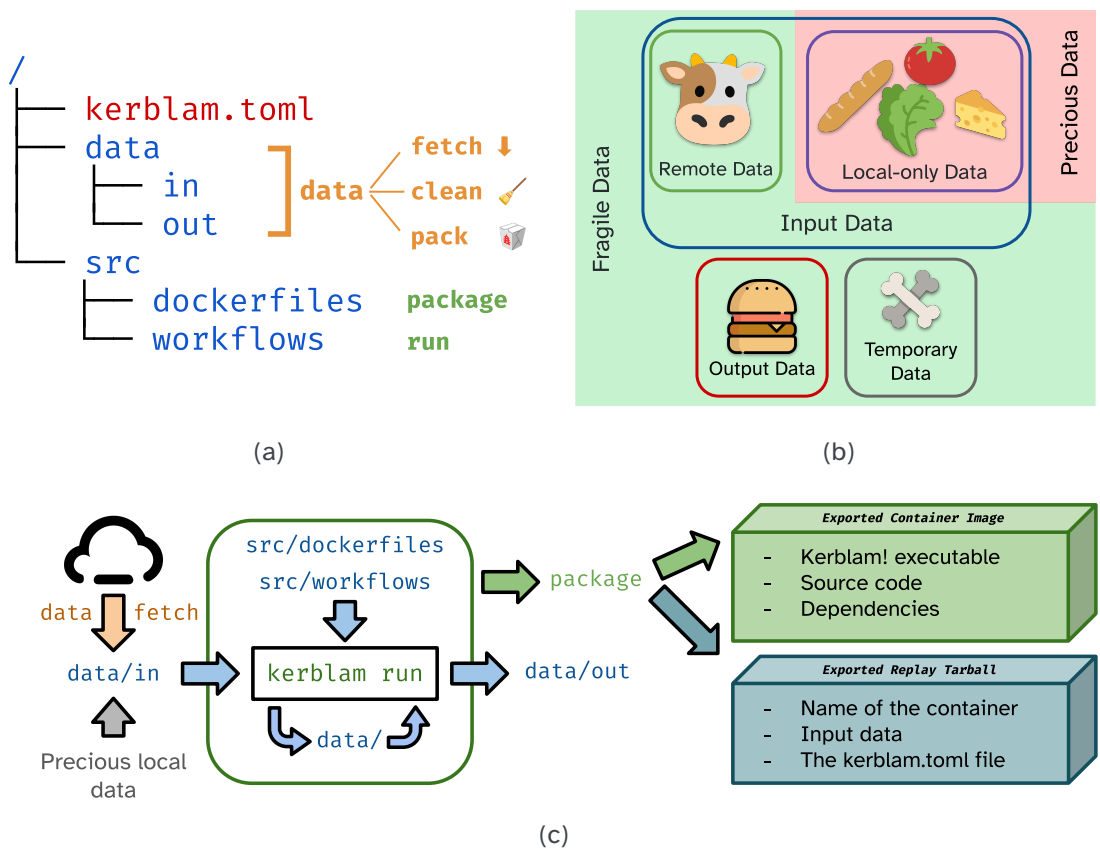

Figure 2: Salient concepts implemented by Kerblam! (a): Basic skeleton of the proposed folder layout for a generic data analysis project associated with relevant Kerblam! commands. Folders are depicted in blue, while files are depicted in red. (b): Data is qualitatively divided into input, output, and temporary data. Input data can be further divided into input data remotely available (i.e., downloadable) and local-only data. The latter is "precious", as it cannot be easily recreated. Other types of data are "fragile", as they may be created again *on the fly*. (c): Overview of a generic Kerblam! workflow.

Kerblam! is a command-line tool written in Rust that incentivizes researchers to use a common, standardized filesystem structure, adopt containerization technologies to perform data analysis, leverage remote file storage when possible, and create and publish readily executable container images to the public to re-run pipelines for reproducibility purposes (see Figure 2c). These aims try to allow and promote the principles described above.

The most basic skeleton of the project layout implemented by Kerblam! can be seen in Figure 2a. The `kerblam.toml` file contains configuration information for Kerblam! and marks the folder as a Kerblam-managed project. Kerblam! provides a number of utility features *out of the box* on projects that adapt to the layout presented in Figure 2a, or any other project structure after proper configuration.

*Data management.* Kerblam! can be used to manage a project's data. Kerblam! automatically distinguishes between input, output, and intermediate data, based on which folder the data files are saved in: the `data` folder contains intermediate data produced during the execution of the workflows, the `data/in` contains input data, and similarly `data/out` contains output data. Furthermore, the user can define in the `kerblam.toml` configuration which input data files can be fetched remotely, and from which endpoint. This allows Kerblam! to both fetch these files upon request (`kerblam fetch`) and to distinguish between remotely available input files and local-only files. Local-only files are deemed "precious", as they cannot be easily recreated. All other data files are "fragile", as they can be deleted without repercussion to save disk space (Figure 2b).

These distinctions between data types enable further functions of Kerblam!. `kerblam data` shows the number and size of files of all types, to quickly check how much disk space is being used by the project. Fragile data can be deleted to save disk space with `kerblam data clean` and precious input data can be exported easily with `kerblam data pack`. `kerblam data pack` can also be used to export output data quickly to be shared with, for example, colleagues.

Allowing Kerblam! to manage the project's data with these tools can offload several chores usually done manually by the experimenter.

*Workflow management.* Kerblam! can manage multiple workflows written for any workflow manager. At its core, Kerblam! can spawn shell subprocesses that then execute the particular workflow manager, potentially one configured by the user. This allows Kerblam! to manage *other* workflow managers, making them transparent to the user and with a single access point.

Kerblam! also can act before and after the workflow manager proper to aid in several tasks. Firstly, Kerblam! can manage workflows in the `src/workflows` folder *as if* they were written in the root of the project. It achieves this by moving the workflow files from said folder to the root of the repository *just before* execution. This allows for slimmer workflows which do not crowd the root of the repository or conflict with each other, thus being more consistent.

Secondly, it allows the concept of *input data profiles*. Data profiles are best explained through an example. Imagine an input file, `input.csv`, containing some data to be analyzed. The experimenter may wish to test the workflows that they have written with a similar, but—say—smaller `test_input.csv`. Kerblam! allows the hot-swapping of these files just before execution of the workflow manager through profiles. By configuring them in the `kerblam.toml` file, the experimenter can execute a workflow manager (with `kerblam run`), specifying a profile: Kerblam! will then swap these two files just before and just after the execution of the workflow to seamlessly use exactly the same workflow but with different input data, in this case for testing purposes.

Kerblam! supports *out of the box* GNU `make` as its workflow manager of choice. Indeed, makefiles can directly be ran through Kerblam! with no further configuration by the user. Any other workflow manager can be used by writing tiny shell wrappers with the proper invocation command. The range of workflow managers supported out-of-the-box by Kerblam! may increase in the future.

*Containerization support.* Containers can be managed directly by Kerblam!. By writing container recipes in `src/dockerfiles`, Kerblam! can automatically execute workflow managers inside the containers, seamlessly mounting data paths and performing other housekeeping tasks before running the container. As already stated, Kerblam! works "above" workflow managers. Therefore, the reader might be questioning the usefulness of a containerization wrapper at the level of Kerblam! if the workflow manager of choice already supports it. This containerization feature is meant to be used when a workflow manager would be inappropriate. For instance, very small analyses might not warrant the increased development overhead to use tools such as CWL. Kerblam! allows even shell scripts to be containerized anyway, making even the tiniest analyses reproducible.

With these capabilities, Kerblam! promotes reproducibility and allows even inexperienced users to perform even the simplest analyses in a reproducible way.

*Pipeline export.* Workflows managed by Kerblam! with an available container can be automatically exported in a reproducible package through `kerblam package`. This creates a preconfigured container image ready to be uploaded to a container registry of choice together with a compressed tarball containing information on how to (automatically) replay the input analysis: the "replay package".

The process automatically strips all unneeded project files, leading to small container images.

The replay package can be inspected manually by a potential examiner, and either re-run manually or through the convenience function `kerblam replay` which recreates the same original project layout, fetches the input container and runs the packaged workflow.

*The Kerblam! analysis flow.* Kerblam! favors a very specific methodology when analyzing data, starting with an empty `git` repository. First, upload the input data in some remote archive (in theory promoting FAIR-er data). Then, configure Kerblam! to download the input data, and write code and workflows to analyze it, potentially in isolated containers or with specific workflow management tools. During development, periodically clean out intermediate and output files to check if the correct execution of the analysis has become dependent on local-only state. Finally, package the results and pipelines into the respective environments and share them with the wider public (e.g., as a GitHub release or in an archive like Zenodo).

We believe that this methodology is simple yet flexible and robust, allowing for high-quality analyses in a wide variety of scenarios.

*Availability.* Kerblam! is a Free and Open Source Software, available on GitHub at [MrHedmad/kerblam](MrHedmad/kerblam). Kerblam! is written in Rust and may be compiled to support both GNU/Linux-flavored operating systems and Mac-OS. Alternatively, GitHub releases provide pre-compiled artifacts for both these operating systems. Support for Windows machines is untested at the time of writing. The full documentation to Kerblam! is available at [kerblam.dev](kerblam.dev). Active support for Kerblam! and its future development are guaranteed for the foreseeable future.

## 5 Conclusions

Structuring data analysis projects is a personal matter, heavily dependent on the preference of the individual(s) who run the analysis. Nevertheless, best practices arise and can be individuated in this fragmented landscape.

With this work, we aimed at providing such guidelines, and included a robust tool to leverage the regularity of such standardized layout. As the proposed layout is, for all intents and purposes, largely arbitrary, Kerblam! can be configured to work with any layout.

Through these and potentially future standardization efforts, tools such as containerization and workflow managers can become more mainstream and even routine, leading to an overall more mature and scientifically rigorous way to analyze data of any kind.

## 6 Code and Data availability

The raw data fetched by the analysis of project templates (e.g., list of fetched repositories, detected frequencies, etc.) are available on Zenodo at the following DOI: [10.5281/zenodo.13627214](). The code for the analysis is available on GitHub at [MrHedmad/ds_project_structure]() and is archived on Zenodo with DOI: [10.5281/zenodo.13627322]().

Kerblam! is available on GitHub at [https://github.com/MrHedmad/kerblam]() and archived at every release in Zenodo at DOI: [10.5281/zenodo.10664806](). Its documentation is available at [https://kerblam.dev/]().

## 7 Author's contributions

Conceptualization: L.V., L.M., and F.A.R.; Software: L.V.; Methodology: L.V. and F.A.R; Resources and Funding Acquisition: L.M.; Writing - Original Draft: L.V., L.M., and F.A.R.; Supervision: L.M. and F.A.R.